\documentclass[
    ,final            
  ]
  {aipproc}

\layoutstyle{6x9}

\def\ga{\mathrel{\mathpalette\fun >}}
\def\fun#1#2{\lower3.6pt\vbox{\baselineskip0pt\lineskip.9pt
  \ialign{$\mathsurround=0pt#1\hfil##\hfil$\crcr#2\crcr\sim\crcr}}}

\begin{document}

\title{Ultra High Energy Cosmic Radiation: Experimental and Theoretical
Status}


\author{G\"unter Sigl}{
  address={APC~\footnote{UMR 7164 (CNRS, Universit\'e Paris 7,
CEA, Observatoire de Paris)} (AstroParticules et Cosmologie),
  11, place Marcelin Berthelot, F-75005 Paris, France and\\
Institut d'Astrophysique de Paris, 98bis Boulevard Arago,
 75014 Paris, France}
}

\begin{abstract}
We give a brief overview of the current experimental and theoretical
status of cosmic rays above $\sim10^{17}\,$eV. We focus on
the role of large scale magnetic fields and on multi-messenger
aspects linking charged cosmic ray with secondary $\gamma-$ray and
neutrino fluxes.
\end{abstract}

\maketitle


\section{Introduction}

High energy cosmic ray (CR) particles are shielded
by Earth's atmosphere and reveal their existence on the
ground only by indirect effects such as ionization and
showers of secondary charged particles covering areas up
to many km$^2$ for the highest energy particles. In fact,
in 1912 Victor Hess discovered CRs by measuring ionization from
a balloon~\cite{hess}, and in 1938 Pierre Auger proved the existence of
extensive air showers (EAS) caused by primary particles
with energies above $10^{15}\,$eV by simultaneously observing
the arrival of secondary particles in Geiger counters many meters
apart~\cite{auger_disc}.

After almost 90 years of research, the origin of cosmic rays
is still an open question, with a degree of uncertainty
increasing with energy~\cite{crbook}: Only below 100 MeV
kinetic energy, where the solar wind shields protons coming
from outside the solar system, the sun must give rise to
the observed proton flux. Above that energy the CR spectrum
exhibits little structure and is approximated
by broken power laws $\propto E^{-\gamma}$:
At the energy $E\simeq4\times 10^{15}\,$eV
called the ``knee'', the flux of particles per area, time, solid angle,
and energy steepens from a power law index $\gamma\simeq2.7$
to one of index $\simeq3.0$. The bulk of the CRs up to at least
that energy is believed to originate within the Milky Way Galaxy,
typically by shock acceleration in supernova remnants.
The spectrum continues with a further steepening to $\gamma\simeq3.3$
at $E\simeq4\times 10^{17}\,$eV, sometimes called the ``second knee''.
There are experimental indications that the chemical composition
changes from light, mostly protons, at the knee to domination by
iron and even heavier nuclei at the second knee~\cite{Hoerandel:2004gv}.
This is in fact expected in any scenario where acceleration and
propagation is due to magnetic fields whose effects only depend
on rigidity, the ratio of charge to rest mass, $Z/A$. This is true
as long as energy losses and interaction effects, which in general depend
on $Z$ and $A$ separately, are small, as is the case in the Galaxy, in
contrast to extra-galactic cosmic ray propagation at ultra-high energy.
Above the so called ``ankle'' or ``dip'' at $E\simeq5\times10^{18}\,$eV, the
spectrum flattens again to a power law of index $\gamma\simeq2.8$.
This latter feature is often interpreted as a cross over from a Galactic
component, which steepens because cosmic rays are not confined by
the Galactic magnetic field any more or because Galactic sources
do not accelerate beyond the ankle, to a harder component of extragalactic
origin. However, the dip at $E\simeq5\times10^{18}\,$eV could also be
explained by pair production by extra-galactic protons,
if the extra-galactic component already starts to dominate below
the ankle, for example, around the second-knee~\cite{Aloisio:2006wv}
at a few times $10^{17}\,$eV. This requires a relatively steep injection
spectrum $\propto E^{-2.6-2.7}$. Below a few times $10^{17}\,$eV
this extra-galactic component would become unobservable at Earth
due to diffusion in extra-galactic magnetic fields (EGMF)~\cite{Lemoine:2004uw}.
In addition, the effective volume-averaged injection spectrum has to become 
flatter somewhere below $\sim10^{18}\,$eV in order to avoid excessive
power going into cosmic rays and to avoid overproduction
of GeV--TeV $\gamma-$rays from $pp$ interactions with the ambient gas.

The low cross-over scenario also requires the dominance of protons around
the dip. Theoretically, this can be achieved either because preferentially 
protons are accelerated or because extended EGMF lead to strong
photo-spallation during propagation~\cite{Sigl:2005md}.
Experimentally, above $\simeq10^{17}\,$eV the chemical composition is basically unknown~\cite{Watson:2004ew}. Around $10^{18}\,$eV the situation is
particularly inconclusive as HiRes~\cite{Abbasi:2004nz} and HiRes-MIA~\cite{Abu-Zayyad:2000ay} data suggest a light (proton dominated) composition, whereas other experiments indicate a heavy composition~\cite{Hoerandel:2004gv}. In any case, the cosmic
ray flux should be extra-galactic at least above the ankle, where a
galactic origin would predict an anisotropy toward the
galactic plane because galactic magnetic fields can no longer
isotropize the cosmic rays. No such anisotropy is seen.
There are also experimental
indications for a chemical composition becoming again lighter above
the ankle, although a significant heavy component is not
excluded and the inferred chemical composition above
$\sim10^{18}\,$eV is sensitive to the model of air shower interactions
and consequently uncertain presently~\cite{Watson:2004ew}.
In addition, should a substantial heavy composition
be experimentally observed up to the highest energies, some sources
would have to be surprisingly nearby, within a few Mpc, otherwise only
low mass spallation products would survive propagation~\cite{er}.
In the following we will restrict our discussion on extra-galactic
ultra-high energy cosmic rays (UHECRs).

Although statistically meaningful information about the UHECR energy
spectrum and arrival direction distribution has been 
accumulated~\cite{Cronin:2004ye}, no
conclusive picture for the nature and distribution of the sources
emerges naturally from the data. There is on the one hand the approximate
isotropic arrival direction distribution~\cite{bm} which indicates that we are
observing a large number of weak or distant sources. On the other hand,
there are also indications which point more towards a small number of
local and therefore bright sources, especially at the highest energies:
First, the AGASA ground array claimed statistically significant multi-plets of
events from the same directions within a few degrees~\cite{Shinozaki:2006kk,bm},
although this is controversial~\cite{Finley:2003ur} and has not been seen so far
by other experiments such as the fluorescence experiment HiRes~\cite{finley}.
The spectrum of this clustered component is $\propto E^{-1.8}$ and thus
much harder than the total spectrum~\cite{Shinozaki:2006kk}.
Second, nucleons above $\simeq70\,$EeV suffer heavy energy losses due to
photo-pion production on the cosmic microwave background (CMB)
--- the Greisen-Zatsepin-Kuzmin (GZK) effect~\cite{gzk} ---
which limits the distance to possible sources to less than
$\simeq100\,$Mpc~\cite{stecker}. This predicts a ``GZK cutoff'', a
drop in the spectrum, whose strength depends on the source distribution
and may even depend on the
part of the sky one is looking at: The ``cutoff'' could be mitigated
in the northern hemisphere where more nearby accelerators related
to the local supercluster can be expected. Apart from the SUGAR array
which was active from 1968 until 1979 in Australia, all UHECR detectors
completed up to the present were situated in the northern hemisphere.
Nevertheless the situation is unclear even there: Whereas a ``cut-off''
is consistent with the few events above $10^{20}\,$eV recorded
by the fluorescence detector HiRes~\cite{hires}, there is a tension
with the 11 events above $10^{20}\,$eV detected by the AGASA ground
array~\cite{agasa}. Still, this could be a combination of statistical
and systematic effects~\cite{DeMarco:2003ig}, especially given the recent
downward revision of the energy normalization in AGASA~\cite{teshima}.
The solution of this problem will have to await more analysis and
more statistics and, in particular, the completion of the Pierre Auger 
project~\cite{Kampert:2006ec}
which will combine the two complementary detection techniques
adopted by the aforementioned experiments and whose southern site
is currently in construction in Argentina.

\section{Role of large scale magnetic fields}

The hunt for UHECR sources is further complicated by the presence
of large scale cosmic magnetic fields which may significantly deflect
charged cosmic rays even at the highest energies, in particular if
sources correlate with high magnetic field regions such as galaxy
clusters. A major issue in UHECR 
propagation studies is, therefore, the strength and distribution of EGMF.
It is known that galaxy clusters harbor magnetic fields of microGauss
strength. Unfortunately, it is poorly known how quickly these fields fall
off with increasing distance from the cluster center. The current data
indicate that $\mu$G strength magnetic fields extend out to at least $\sim 1$
Mpc~\cite{clarke}
and possibly to larger distances~\cite{johnston,Govoni:2006fs}.
At distances above $1\,$Mpc from a cluster core, however, probing the
magnetic fields becomes extremely difficult because the Faraday
Rotation Measure loses sensitivity in low density
regions. Furthermore, the intracluster magnetic field topology is
also poorly known, although the situation will likely improve
in the future, for example with the advent of powerful radio astronomical
instruments such as the square kilometer array.

One possibility in the meantime is to adopt large scale structure simulations (LSS) which include magnetic fields. However, different models for these
fields tend to give different predictions for UHECR deflection,
as the comparison between Refs.~\cite{lss-protons,sme-proc}
and Ref.~\cite{dolag} shows. In Ref.~\cite{lss-protons}, the authors
use magnetic fields derived from a cosmological
LSS with magnetic fields generated at the shocks that
form during LSS formation, whereas in Ref.~\cite{sme-proc} and
Ref.~\cite{dolag} fields of ``primordial'' origin have been
considered. While the different models for initial magnetic seed fields
produce different large scale magnetic field distributions and,
therefore, lead to different predictions for UHECR deflection,
there is still a significant discrepancy between
Ref.~\cite{lss-protons,sme-proc} and  Ref.~\cite{dolag}, hinting
that other technical reasons may play a role here. In the more
extended fields from the simulations
of Refs.~\cite{lss-protons,sme-proc} deflection of protons up to
$10^{20}\,$eV can be up to tens of degrees, whereas deflections
in the simulations of Ref.~\cite{dolag} are typically below a degree.

We recall that since acceleration is rigidity dependent, at the
acceleration sites the highest energy cosmic ray flux is likely
dominated by heavy nuclei. If this is indeed the case, it is interesting
to point out that even in the EGMF scenario of Ref.~\cite{dolag},
deflections could be considerable and may not allow particle astronomy
along many lines of sight: The distribution of deflection angles in
Ref.~\cite{dolag} shows that deflections of protons above
$4\times10^{19}\,$eV of $\ga1^\circ$ cover a considerable fraction of
the sky. Suppression of deflection along typical lines of sight by
small filling factors of deflectors is thus unimportant in this case.
The deflection angle of any nucleus at a given energy passing through
such areas will therefore be roughly proportional to its charge as
long as energy loss lengths are larger than a few tens of
Mpc~\cite{bils}. Deflection angles of $\sim20^\circ$ at
$\sim4\times10^{19}\,$eV should thus be the rule for iron nuclei. In
contrast to the contribution of our Galaxy to deflection which can be
of comparable size but may be corrected for within sufficiently
detailed models of the galactic field, the extra-galactic contribution
would be stochastic. Statistical methods are therefore likely to be
necessary to learn about UHECR source distributions and
characteristics as well as EGMF. For example, a suppressed UHECR arrival 
direction auto-correlation function at degree scales, rather than
pointing to a high source density, could be a signature of
extended EGMF~\cite{lss-protons}.

Finally, EGMF can considerably increase the path-length of UHECR
propagation and thus spectra, especially from individual sources,
as well as the chemical composition observed at Earth~\cite{spec-comp}.

\section{Multi-messenger approach: Charged primary cosmic rays and
secondary gamma-rays and neutrinos}

\begin{figure}[h!]
\includegraphics[height=0.5\textwidth]{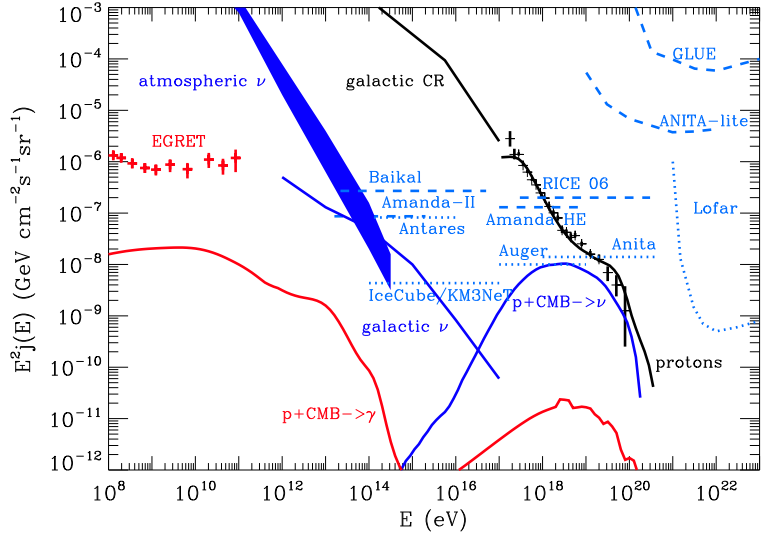}
\caption{
Model fluxes (multiplied by the squared energy) compared to experimental
data, limits and sensitivities. Primary cosmic ray fluxes (data and a model,
see text) are shown in black, the secondary  $\gamma-$ray flux expected from proton interactions with the CMB and infrared background
in red and the "guaranteed" neutrino fluxes per neutrino flavor in blue: 
atmospheric neutrinos, galactic neutrinos resulting from cosmic ray interactions with matter in our Galaxy, and "GZK'' neutrinos resulting from cosmic ray interaction with the CMB and infrared background. These secondary fluxes depend to some extent on the distribution of the (unknown) primary cosmic ray sources for which active galaxies were assumed above $10^{17}\,$eV. The flux of atmospheric neutrinos has been measured by underground detectors and AMANDA. Also shown are existing upper limits and future sensitivities to diffuse neutrino fluxes from various experiments (dashed and dotted light blue lines, respectively)~\cite{nu_review}, assuming the Standard Model neutrino-nucleon
cross section extrapolated to the relevant energies. The maximum possible neutrino flux would be given by horizontally extrapolating the diffuse $\gamma-$ray background observed by EGRET~\cite{Strong:2003ex}.}
\label{fig1}
\end{figure}

The physics and astrophysics of UHECRs are also intimately linked with
the emerging field of neutrino astronomy~\cite{nu_review} as well as
with the already well established field of $\gamma-$ray 
astronomy~\cite{gammarev}. Indeed, all
scenarios of UHECR origin, including the top-down models, are severely
constrained by neutrino and $\gamma-$ray observations and limits.
In turn, this linkage has important consequences for theoretical
predictions of fluxes of extragalactic neutrinos above about a TeV
whose detection is a major goal of next-generation
neutrino telescopes: If these neutrinos are
produced as secondaries of protons accelerated in astrophysical
sources and if these protons are not absorbed in the sources,
but rather contribute to the UHECR flux observed, then
the energy content in the neutrino flux can not be higher
than the one in UHECRs, leading to the so called Waxman-Bahcall
bound for transparent sources with soft acceleration
spectra~\cite{wb-bound,mpr}.
If one of these assumptions does not apply, such as for acceleration
sources with injection spectra harder than $E^{-2}$ and/or opaque
to nucleons, or in the top-down scenarios where X particle decays
produce much fewer nucleons than $\gamma-$rays and neutrinos,
the Waxman-Bahcall bound does not apply, but the neutrino
flux is still constrained by the observed diffuse $\gamma-$ray
flux in the GeV range.

Fig.~\ref{fig1} provides a sketch of "realistic'' cosmic
ray, $\gamma-$ray, and neutrino flux predictions in comparison
with experimental observations, limits, and sensitivities. It
shows a theoretical scenario in which
extra-galactic cosmic ray sources roughly evolving as quasars
inject a spectrum $\propto E^{-2.6}$ of dominantly protons down to 
$\sim10^{17}\,$eV where a cross-over to galactic cosmic rays occurs~\cite{Aloisio:2006wv}. The "cosmogenic'' neutrino flux
produced by protons interacting with the low energy photon
background can in principle be used to test these assumptions
on which it depends considerably~\cite{Stanev:2006su}.

\begin{figure}[h!]
\includegraphics[width=0.6\textwidth]{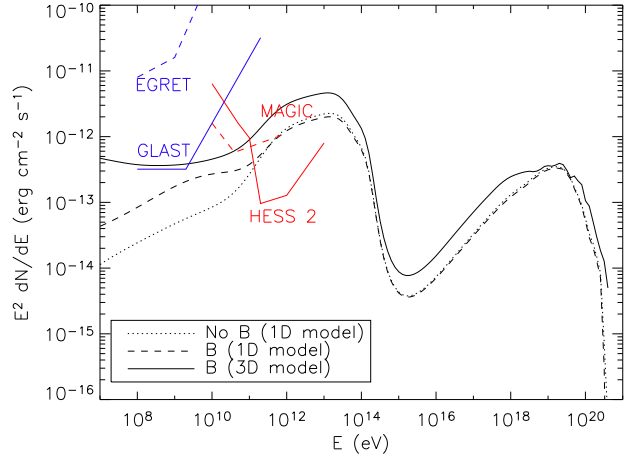}
\caption{Differential $\gamma$-ray fluxes (multiplied by squared energy) from electromagnetic cascades including synchrotron radiation, initiated by
photo-pion and pair production by protons injected with an $E^{-2.7}$ spectrum
(not shown) by a source at distance $d=20\,$Mpc.
We assume the source contributes a 
fraction $\simeq 0.2$ to the total UHECR flux, corresponding to a proton luminosity $\simeq 4 \times 10^{42}\,$erg s$^{-1}$ above
$10^{19}\,$eV. A structured magnetic field of 0.1--1$\mu$G extends a few Mpc around the source in case of the 1D and 3D simulations which take into
account synchrotron radiation of electrons and positrons. The 1D model
neglects proton deflection whereas the 3D simulation follows 3-dimensional
proton trajectories. The latter case implies that the
fluxes shown here would be extended over $\sim5^\circ(20\,{\rm Mpc}/d)$
on the sky.
The fluxes are comparable to sensitivities of space-based $\gamma-$ray
(blue) and ground-based imaging air \v{C}erenkov detectors (red).}
\label{fig2}
\end{figure}

3-dimensional propagation in structured large scale magnetic fields
also has considerable influence on secondary $\gamma-$ray and neutrino
fluxes. Fig.~\ref{fig2} demonstrates how magnetic fields of
micro-Gauss strength surrounding a UHECR source, for example in
a galaxy cluster, can influence the secondary GeV-TeV $\gamma-$ray fluxes 
produced by electromagnetic cascades initiated by UHECR interactions
with the CMB and infrared background. This is the result of simulations
with our public code CRPropa~\cite{crpropa}, discussed in
Ref.~\cite{armengaud,Armengaud:2006fx}. For the
steep injection spectrum $\propto E^{-2.7}$ assumed in Fig.~\ref{fig2},
the photon flux below a TeV is dominated by synchrotron radiation from
electron/positron pairs produced by protons around the ankle. As a
consequence, it depends considerably on strength and extension of the magnetic 
field. We note that apparently time-variable fluxes of order 
$(3-10)\times10^{-13}\,{\rm erg}\,{\rm cm}^{-2}\,{\rm s}^{-1}$ above
730 GeV, concentrated within $\sim0.2^\circ$ of M87 in the Virgo cluster
at $\sim16\,$Mpc, have been seen by HEGRA and HESS~\cite{Beilicke:2005kd}.
A flux of $\sim5\times10^{-12}\,{\rm erg}\,{\rm cm}^{-2}\,{\rm s}^{-1}$
above 1 TeV, extended over a few degrees around the active galactic nucleus
1ES 1959+650 at $\simeq200\,$Mpc, has recently be seen by 
MAGIC~\cite{Aliu:2005jv}.
The spectra tend to be considerably steeper than $E^{-2}$.
Since the fluxes are comparable to the predictions in 
Fig.~\ref{fig2}, these observations indeed start to constrained
the contribution of sources immersed in galaxy clusters to the
UHECR flux.

\begin{theacknowledgments}
I would like to thank Eric Armengaud and Francesco Miniati for recent
collaborations and Dominik Els\"asser and Ralf Wischnewski for useful
information. This work was
partly supported by the European Union under the ILIAS project,
contract No.~RII3-CT-2004-506222.
\end{theacknowledgments}



\bibliographystyle{aipprocl} 

\end{document}